\documentclass[10pt]{emulateapj}
\usepackage{psfig}
\usepackage{graphicx}

\shorttitle{X-ray Binaries and Young Clusters}
\shortauthors{Sepinsky, Kalogera, \& Belczynski}

\begin{document}

\title{Are Supernova Kicks Responsible for X-ray Binary Ejection from Young Clusters?} 

\author{J.\ Sepinsky\altaffilmark{a}\footnote{NASA GSRP Fellow},  V.\ Kalogera\altaffilmark{a}, and 
K.\ Belczynski\altaffilmark{b}\footnote{Tombaugh Postdoctoral Fellow}}

\affil{$^{a}$ Department of Physics and Astronomy, Northwestern University, Evanston, IL 60208\\
$^{b}$ Department of Astronomy, New Mexico State University, Las Cruces, NM 88003\\
j-sepinsky@northwestern.edu, vicky@northwestern.edu, belczynski@northwestern.edu\\}

\begin{abstract}

Recent {\it Chandra} observations of interacting and starburst galaxies have led us to investigate
the apparent correlation between the positions of young star clusters and {\em Chandra} point
sources.  Assumed to be X-ray binaries (XRBs), these point sources do not seem to coincide with the
massive ($\sim$10$^5$\,M$_\sun$), young ($1-50$\,Myr) stellar clusters that can easily form systems
capable of such emission.  We use a sophisticated binary evolution and population synthesis code
({\it StarTrack}) and a simplified cluster model to track both the X-ray luminosity and position of
XRBs as a function of time.  These binaries are born within the cluster potential with
self-consistent positions and velocities and we show that a large fraction ($\simeq$70\%) can be
ejected from the parent due to supernova explosions and associated systemic velocities. For
brighter sources and cluster masses below $\sim$10$^{6}$\,M$_{\sun}$, we find that the average
number of bright XRBs per cluster remains near or below unity, consistent with current
observations. 

\end{abstract}

\keywords{galaxies: star clusters -- methods: 
statistical -- X-rays: binaries} 

\section{Introduction}

One of the interesting developments made possible with {\it Chandra} observations is the abundance
of extragalactic point sources.  The large observed samples now available promise to improve our
understanding of their formation and interaction with their host stellar environments. In galaxies
with young stellar populations, it is generally thought that bright point
X-ray sources are young X-ray Binaries (XRBs) \citep{KIL02, FAB01, FW05}, associated 
with the large amount of ongoing star
formation.  This interpretation is mainly based on the measured X-ray luminosity and the 
spectral and temporal variability characteristics of the point sources.  For a detailed 
review of the formation and evolution of XRBs see \citet{LVDK05}, and for a recent 
discussion of the temporal properties see \citet{SES03}.  Optical and 
infrared observations, most prominently with the {\it Hubble} Space
Telescope, reveal massive, young clusters, often referred to as super star clusters. They range in
mass from $\sim$10$^4$\,M$_\sun$ to $\sim$10$^7$\,M$_\sun$ \citep{SG01, HAR01}, and in age from
just $\sim$1\,Myr to $\sim$50\,Myr, with the majority at $\sim10-20$\,Myr (mostly due to
photometric selection).  These clusters are thought to be young analogs of globular clusters and
may be responsible for most of the massive stars in the field of their host galaxies \citep{TRE01}. 
Thus, one may expect a concetration of XRBs in or near these clusters. 

Recently, {\citet{KAA04}; hereafter K04} have studied 3 starburst galaxies (M82, NGC1569, and
NGC5253) each containing a significant number of young star clusters and point X-ray sources. 
Indeed, they do find a statistically significant relationship between the two types of objects: XRBs
are preferentially found within distances of $\sim$30$-$100\,pc from their nearest cluster, but
there is a clear lack of XRBs found coincident with the clusters.  There are obvious observational
biases, the most important being that the true parent cluster is unknown; the distances quoted are
only those to the {\it nearest} cluster, not necessarily the parent cluster.  Still, the XRB spatial
distribution relative to the clusters and their association seems significant and characteristic of
a non-random sample distribution (see \S\,3 of K04 for the relevant statistical analysis). It is
worth noting that similar results have been found in observations of the Antennae \citep{ZEZ02}. 

In this {\em Letter}, our goal is to model in a self-consistent manner the population of binaries in
the cluster potential (assumed static for simplicity). We track the kinematic evolution of compact
object binaries in the absence of dynamical interactions and we follow their X-ray luminosity
($L_{X}$) evolution.  We focus on two specific, testable points of comparison between the published
observations and our calculations: the average number of XRBs per cluster, and the median distance
of XRBs from their parent cluster (or nearest cluster in the observations). For point sources
brighter than $\sim$10$^{36}$\,erg\,s$^{-1}$, K04 have shown that the median distance from a cluster
is $\sim$100\,pc, with an average of $\lesssim$1 XRB per cluster (See Table 1 in K04 for the
specifics regarding each of the three galaxies considered).  Here we show that these two quantities
can be calculated theoretically and the results appear consistent with the observations for the
range of cluster masses and ages relevant to the clusters observed, when considering only the
supernova kicks imparted to XRBs and their motion in cluster potentials.  Guided by the $L_{X}$ 
sensitivity limits of the observations, we focus our analysis on XRBs with 
$L_{X}\geq$5$\times$10$^{35}$\,ergs\,s$^{-1}$, although the formation and evolution of XRBs to 
lower $L_{X}$ ranges is included in our models.

In \S\,2 we describe the model methods used and how they are applied in our simulations. In \S\,3 we
describe our main results and compare them to the observations presented by K04. We discuss our
conclusions in \S\,4. 

\section{Theoretical Modeling}

To generate the necessary stellar populations for the modeled clusters, we use the population
synthesis program {\it StarTrack} (developed by \citet{BKB02}; Belczynski et al.\ 2004, to be
submitted).  We generate and evolve a population of binaries under a given set of conditions, such
as the initial mass function (IMF), supernova kick distribution, common envelope efficiency, etc. 
With the resultant evolutionary parameters of the binaries at the time of the compact-object
(neutron star or black hole) formation, we place them in a cluster potential and track their motion
and X-ray luminosity as a function of time. In so doing, we ultimately generate a complete
evolutionary picture of the X-ray binaries in association with their parent cluster. As noted
already, we do not account for any stellar interactions in these young clusters, as our goal is to
examine whether supernova kicks alone can account for the observed spatial distribution of XRBs
relative to their parent clusters. 

{\it StarTrack} is a sophisticated Monte Carlo population synthesis code that has been recently
updated to carefully account for binary mass-transfer phases and $L_{X}$ calculation. We account for
various phases of mass and angular momentum losses, and have implemented an integrated tidal
evolution method that is calibrated against observations of Galactic high-mass XRBs and of
circularization in open clusters. Some key features for this investigation include: (i) the
determination of the post-core-collapse systemic velocity for compact object binaries, and (ii) the
detailed calculation of the mass transfer rate between binary components, calibrated against
calculations with a stellar evolution code.  Systemic velocities are a key to the proper
determination of the orbital trajectory, which is one of the primary concerns for this work.  Also,
given the sensitivity of the observations to $L_{X}$, the mass transfer rate becomes a critical
factor in determining whether or not a given XRB is relevant to the K04 observations at any point in
its lifetime. For the $L_{X}$ determination we further apply a bolometric correction to the
theoretical value, to account for {\em Chandra}'s sensitivity band and the typical XRB spectra. This
bolometric correction is dependent on the system parameters (neutron star or black hole accretor,
wind accretion or Roche-lobe overflow) and assumptions about the typical spectra of different
sources, derived empirically from Galactic observations (Maccarone 2003, private communication;
\citet{PZDM04}). Specifically for wind accretors we adopt 0.15 and 0.7, for disk persistent sources
and transient sources at outburst we adopt 0.5 and 0.7 (for neutron stars and black holes,
respectively).
 
It is well known that population synthesis calculations require a significant number of input
parameters. Since our main goal in this {\em Letter} is a proof-of-principle study, it is not
necessary to fully explore the parameter space. Instead we choose to consider a reference model with
parameter assumptions that are considered typical for binary evolution calculations (see model A in
\citet{BKB02}). We also consider a small set of other models where we vary the initial mass function
of binary primaries, as this most significantly affects the relative contribution of XRBs with
neutron stars and black holes that acquire systematically different systemic velocities. 

The systemic velocities are determined by the natal kicks imparted during the formation of the 
compact object.  Neutron star natal kick magnitudes are drawn from the distribution derived from 
\citet{ACC02}, based on current pulsar kinematics.  It consists of two 
Maxwellians with $\sigma$'s 90 and 500, with a relative weight of 2:3, respectively.  Black hole natal kicks are linearlly 
scaled from Neutron star 
kicks based on their mass relative to a typical Neutron star mass of 1.44\,M$_\sun$.

For the orbital evolution we use the calculated post-core-collapse systemic velocities of XRB
progenitors and combine them consistently with a Plummer model for the gravitational potential of
the model cluster with a given half-mass radius, $R_{1/2}$. To determine the spatial distribution of
the XRB progenitors right after compact-object formation (which generally occurs immediately prior
to the X-ray phase), we assume that the number density of stars is proportional to the mass density.
Initial systemic velocities consistent with the Plummer potential are also generated (see
\citet{AAR74}).  We then apply the calculated post-core-collapse systemic velocities randomly
oriented with respect to the initial cluster velocities. We follow the motion of the binary as a
function of time and correlate position with the X-ray luminosity evolution calculated with the
binary evolution code.

The number of binaries modeled for a given cluster is directly proportional to the mass of that
cluster once the IMF index, mass-ratio-distribution paramters, and binary fraction are chosen.  We
adopt a flat mass-ratio distribution and a binary fraction equal to unity in order to represent an
upper limit on the number of binaries, and therefore on the total number of X-ray sources.  We
consider cluster masses in the range $10^4$\,M$_\sun - 10^6$\,M$_\sun$, with IMF indices of 2.35
and 2.7.  The specific parameters of our simulations are shown in Table~1.

\begin{deluxetable}{cccccc}
\tabletypesize{\scriptsize}
\tablecaption{Parameters for model runs}
\tablewidth{0pt}
\tablehead{
\colhead{Model} & \colhead{\# of} & \colhead{Mass} &\colhead{IMF} & \colhead{R$_{1/2}$}\\
& \colhead{MC runs} & \colhead{(M$_\sun$)} & \colhead{index} & \colhead{(pc)}
}
\startdata
A & 1000 & 5$\times$10$^4$ & 2.35 & 10\\
B & 1000 & 5$\times$10$^4$ & 2.7  & 10\\
C & 100  & 5$\times$10$^5$ & 2.35 & 10\\
D & 100  & 5$\times$10$^5$ & 2.7  & 10\\
E & 7    & 5$\times$10$^6$ & 2.35 & 10\\
J & 500  & 5$\times$10$^4$ & 2.35 & 1 
\enddata
\end{deluxetable}

It is interesting to note that although orbits are calculated typically for few to several hundred
Myr, binaries are X-ray sources for only a small part of their orbit and they are {\em bright}
(i.e., Lx $\geq$5$\times$10$^{35}$\,erg\,s$^{-1}$) X-ray sources for an even smaller part.  Each
system is evolved individually in a static cluster potential, and thus no interactions or cluster
evolution is allowed in the present analysis since our goal is to examine the effect of the
supernova kicks. This may not be a well justified assumption in general. Nevertheless, the clusters
relevant to our study are very young (few Myrs to $\sim10-20$\,Myr), so significant cluster
evolution is not expected, except for possibly in the most massive and most compact (small half-mass
radius) clusters. 

We note that statistical effects play a significant role, especially in the low mass
($\sim$10$^4$M$_\sun$) clusters.  Typically, no more than one XRB is bright enough to be seen in
these clusters and the position can vary significantly across the cluster for each different
simulation. Therefore, we consider a large number of Monte Carlo realizations for each parameter set
(cluster mass, half-mass radius, and IMF index).  The lower mass clusters have the smallest number
of initial binaries and hence require the most realizations. We chose the number of realizations for
each cluster mass so that our results averaged over the many realizations remained unaffected at the
5\% level.

\section{Results \& Discussion}

Our calculations have yielded a wealth of results.  Knowing both the trajectory and X-ray luminosity
of such a wide range of objects can help us understand XRB formation in young clusters both
statistically and in a system-by-system sense.  Here, we focus on the statistical results where the
clusters are investigated as a grand average; this is more appropriate when comparing with large
populations of clusters (as analyzed in K04).  Specifically, we focus on the statistical averages of
the two quantites quoted observationally in K04: the median distance of XRBs from the nearest
(parent in the models) cluster, and the mean number of bright XRBs per cluster (within 1000\,pc).

In Figure~1 (top) we plot the model average number of XRBs per cluster as a function of distance
from the parent clusters, each of 5$\times$10$^4$ M$_\sun$ and for a variety of cluster ages.  These
ages are within the range of estimates for the observed clusters. To take into account the
uncertainties in the age estimates (typically a few Myr), we use an age-snapshot method, based on
which we determine the average number of XBRs as a function of radius for a specific ``instant" in
time, and then average these results over each consecutive ``instant" within the cluster age
estimate and its error. 

\begin{figure}
\begin{center}
\includegraphics[width=6cm,height=17cm]{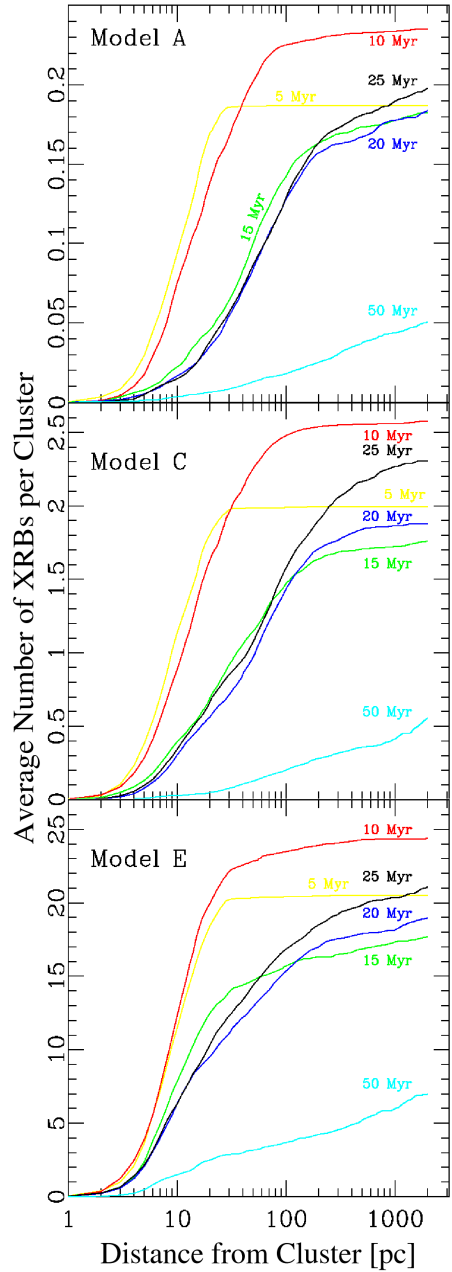}
\caption{Average number of XRBs seen within a given distance from its 
parent cluster for specific cluster ages listed shown.  See Table~1 for the specifications of 
the model parameters.  All data shown here is calculated with an L$_X$ cutoff of 5$\times$10$^{35}$\,erg\,s$^{-1}$.}
\end{center}
\end{figure}

It is evident from Figure~1 that the XRB spatial distributions have a dramatic time dependence. For
``young'' clusters, the average XRB number per cluster rises to a maximum rapidly and very few XRBs
are found at large distances. This is primarily because even the unbound XRBs have not had enough
time to move away from their parent clusters. The median systemic velocity of the XRB systems is
$\sim$2$-$3\,pc\,Myr$^{-1}$, which limits the distance any XRB can reach. For older clusters, the
average XRB number exhibits a fairly slow increase with distance, up to 2\,kpc and sometimes 
beyond.  This 
can potentially create a pollution effect and lead to difficulty in identifying the true 
parent cluster in observations.

It is also evident that at certain ages XRBs are distinctly more numerous than at others.  For
example, in Figure~1, the 5$\times$10$^4$\,M$_\sun$ (top) clusters with
$L_{X}\geq$5$\times$10$^{35}$\,erg\,s$^{-1}$ show more XRBs at 10\,Myr than at any other time in the
clusters' evolution.  We also find that this peak age is dependent on the $L_{X}$ cut-off.  Fully
exploring these dependencies could allow us to derive general conclusions about XRB populations
dependent only on the average ages of the young cluster population.

In Figure~1 (middle and bottom) we present our results for clusters of 5$\times$10$^5$ and
5$\times$10$^6$\,M$_\sun$, respectively.  Note that the behavior is similar for all masses, except
that the average number of XRBs at a given radius scales with the mass of the cluster almost
lineraly.  This is due to the direct relationship between the number of binary systems modeled and
the cluster mass. 

\begin{deluxetable*}{ccccccccccccc}
\tabletypesize{\footnotesize}
\tablecaption{Mean XRB Number ($\overline{N}_{XRB}$) and Median\tablenotemark{a} XRB distance from cluster center (R$_{median}$)} \tablewidth{0pt}
\tablecolumns{13}
\tablehead{
\colhead{} & \multicolumn{2}{c}{5 Myr} & \multicolumn{2}{c}{10 Myr} & \multicolumn{2}{c}{15 Myr} & 
\multicolumn{2}{c}{20 Myr} & \multicolumn{2}{c}{25 Myr} & \multicolumn{2}{c}{50 Myr} \\
\colhead{Model} & \colhead{$\overline{N}_{XRB}$} & \colhead{R$_{median}$} & 
\colhead{$\overline{N}_{XRB}$} & \colhead{R$_{median}$} & 
\colhead{$\overline{N}_{XRB}$} & \colhead{R$_{median}$} &
\colhead{$\overline{N}_{XRB}$} & \colhead{R$_{median}$} &
\colhead{$\overline{N}_{XRB}$} & \colhead{R$_{median}$} &
\colhead{$\overline{N}_{XRB}$} & \colhead{R$_{median}$} \\
 & & (pc) & & (pc) & & (pc) & & (pc) & & (pc) & & (pc)
}
\startdata
A & 0.19 & 10.5 & 0.23 & 17.5 & 0.18 & 44.5 & 0.18 & 54.5 & 0.19 & 58.5 & 0.04 & 146.5 \\
B & 0.14 & 10.5 & 0.18 & 17.5 & 0.12 & 46.5 & 0.12 & 67.5 & 0.16 & 76.5 & 0.04 & 104.5 \\
C & 2.00 &  9.5 & 2.55 & 15.5 & 1.72 & 27.5 & 1.87 & 46.5 & 2.26 & 51.5 & 0.41 &  80.5 \\
E & 20.5 & 16.5 & 24.3 & 17.5 & 17.3 & 18.5 & 18.1 & 24.5 & 20.4 & 26.5 & 6.00 &  29.5 \\
J & 0.18 &  1.5 & 0.25 &  2.5 & 0.18 & 13.5 & 0.16 & 31.5 & 0.21 & 40.5 & 0.04 &  54.5  
\enddata
\tablenotetext{a}{Only XRBs within 1000\,pc are used to calculate the values listed here in order to 
compare with K04}
\end{deluxetable*}

We calculate the median distance and mean number of XRBs with
$L_{X}\geq$5$\times$10$^{35}$\,erg\,s$^{-1}$ within 1000\,pc (Table~2), in order to compare
appropriately with K04. 

{\em Mean number of XRBs per cluster:} We find the theoretical mean XRB number per cluster to vary
significantly from $\sim 0.1$ to $\sim 10$, depending on the cluster mass.  Therefore, it is
possible to reproduce the results in K04 by taking contributions from a number of clusters of
different masses.  Two of the three galaxies discussed in K04 (M82 and NGC~5253) have a mean number
of observed XRBs of $\sim 1$ per cluster, while NGC~1569 seems to have a very small number of XRBs
(only $\simeq 0.25$ per cluster). This difference would point towards NGC~1569 having, on average,
smaller-mass clusters, even though outliers at high masses can still exist. A difficulty in the
comparison arises because the properties of the clusters in these galaxies are difficult to
determine orbservationally.  Those with measured masses are skewed to higher masses
($\gtrsim$1$\times$10$^5$\,M$_\sun$) and younger ages ($\lesssim$15\,Myr) simply because they are
selected photometrically (Gallagher 2004, private communication).  Therefore, developing a proper
theoretical cluster distribution for comparison is rather challenging without further observational
studies of the cluster populations. 
 
{\em Median distance of XRBs from the cluster:} Our results (Table 2) indicate a strong dependence
of the median XRB distance on the age and a moderate dependence on the cluster mass. For clusters
with a half-mass radius of $10$\,pc and masses $\lesssim5\times10^5$\,M$_\sun$, median distances
reach values of $30-100$\,pc (similar to those observed) at ages of $15$\,Myr and older. Only very
massive clusters of $\sim 5\times10^{6}$\,M$_{\sun}$ reach such distances later at $\sim 50$\,Myr.
These ages and moderate masses are consistent with the current observational estimates, although
massive and older clusters are also present in the photometrically selected clusters in K04
(Gallagher 2004, private communication). 

It should also be noted that, for the highest cluster mass we consider (5$\times$10$^6$ M$_\sun$),
even the oldest clusters seem to show more binaries than what is observed. This clearly implies that
starbursts are not dominated by such massive clusters, and this is not surprising. However, these
more massive clusters may also be affected by dynamical cluster evolution and stellar interactions
leading more binary disruptions and ejections. Thus we would expect the average number of XRBs per
cluster to decrease at all ages and distances.

It should be noted that we have assumed a binary fraction of unity, and therefore the mean XRB
numbers could be overestimated.  This is true also because projection effects have not been taken
into account, and our numbers represent the radial distance the XRBs have traveled.  Also, we note
that changes in the power law IMF index of the cluster produce noticeable, but largely insignificant
changes in the cluster profiles. For example, changing the IMF index from 2.35 to 2.7 decreases the
average number of binaries at or about the 10\% level for each timestep.  This effect may become
more important, especially for very steep IMFs, such as those for clusters proposed by \citet{KW03}
where the index can go as high as 3.2.  And last, changes in the half-mass radius of the cluster
dramatically change the median XRB distance for a given mass.  Very small values (model J in
Table~2) tend to limit XRBs ejection, as the potential is deeper.  In these tight clusters, it is
likely that dynamics will play a non-negligible role, depending on their age.

\section{Conclusions}

With detailed population simulations of XRBs and a simple treatment of gravitational potentials of
young clusters we have shown that the significantly low XRB numbers per cluster observed in
starbursts can be explained as being largely due to supernova kicks imparted to XRBs at
compact-object formation that lead to XRB ejection from the cluster potential, as heuristically
suggested by observational studies (K04; \citet{PZDM04}). Derived XRB median distances are also
consistent with current estimates of cluster masses and ages, although a more direct comparison
requires more detailed observational constraints of the cluster properties.

This work opens many possible avenues in which to continue this study, some of which include an in
depth look into the systematics generated by our stellar evolution code, such as how our results
change with a broader range of masses and IMFs, as well as additional stellar evolution parameters
such as the common envelope efficiency.  We also intend to look at the detailed populations created,
and search for specific correlations between types of XRBs, their ages, and positions in the
clusters.  Still further, we have largely ignored the low luminosty XRBs in this analysis.  This
population may indeed be detectable, if present in large enough quantities, as diffuse emission. 
And lastly, it is possible that for the more massive, compact, and older clusters, dynamics play a
non-negligible role in the XRB evolution.  We hope to extend our modeling to include dynamical
considerations such as this in the near future.

\acknowledgments

We are grateful to J.\ Gallagher, T.\ Maccarone, and P.\ Kaaret for useful discussions. The work is 
partially a Packard Foundation fellowship and a NASA Chandra Award to V.\ Kalogera.

\end{document}